\documentclass[%
 reprint,
 amsmath,amssymb,
 aps, showpacs
]{revtex4-2}

\usepackage{graphicx}
\usepackage{dcolumn}
\usepackage{bm}
\usepackage{verbatim}
\usepackage{array}
\usepackage{amsmath}
\usepackage{braket}
\usepackage{lineno}
\usepackage[bookmarksnumbered, pdfstartview=FitH,colorlinks,urlcolor=blue, citecolor=blue,linkcolor=blue,]{hyperref}
\usepackage{setspace}
\usepackage{subfigure}

\begin{document}

\preprint{APS/123-QED}

\title{Searches for New Physics Beyond the Standard Model in Hyperon Sector}

\author{Jianyu Zhang$^{1,2}$}
\email{zhangjianyu@ucas.ac.cn}

\author{Jinlin Fu$^{1}$}
\email{jinlin.fu@ucas.ac.cn}

\author{Hai-Bo Li$^{1,3}$}
\email{lihb@ihep.ac.cn}

\affiliation{$^{1}$School of Physical Sciences, University of Chinese Academy of Sciences, Beijing 100049, China}
\affiliation{$^{2}$National Centre for Nuclear Research, Pasteura 7, 02-093 Warsaw, Poland}
\affiliation{$^{3}$Institute of High Energy Physics, Chinese Academy of Sciences, Beijing 100049, China}

\begin{abstract}
Hyperon physics offers a distinctive laboratory for probing the intensity frontier and searching for physics beyond the Standard Model.
This review summarizes recent results from the BESIII experiment, including pioneering studies of dark baryons, massless BSM particles, and invisible decay modes, together with investigations of baryon- and lepton-number violation.
A central highlight is the determination of the $\Lambda$ electric dipole moment using quantum-entangled hyperon–antihyperon pairs, achieving a sensitivity three orders of magnitude beyond previous limits.
These measurements provide world-leading constraints on new physics scenarios and establish a robust foundation for next-generation precision studies.
By integrating experimental progress with theoretical developments and future facility prospects, this review emphasizes the critical role of hyperon probes in testing the fundamental laws of nature.
\end{abstract}



\maketitle

\section{Introduction}

The Standard Model (SM) of particle physics has achieved remarkable success in describing fundamental interactions, yet several profound questions remain unresolved, including the nature of dark matter and the origin of the cosmic matter--antimatter asymmetry~\cite{Sakharov:1967dj}. 
These open issues indicate that the SM should be regarded as an effective theory, motivating high-precision searches for physics beyond the Standard Model (BSM) at the intensity frontier. 
Hyperons—baryons containing strange quarks—play a central role in this program. 
As spin-$1/2$ systems undergoing parity-violating weak decays, hyperons are intrinsically ``self-analyzing,'' enabling direct access to polarization observables and decay asymmetry parameters. 
The Beijing Spectrometer~III (BESIII) experiment at the BEPCII collider has effectively become a high-luminosity hyperon source, exploiting a data sample of approximately $10^{10}$ $J/\psi$ events~\cite{BESIII:2009fln, BESIII:2021cxx} to produce large, clean, and quantum-entangled hyperon--antihyperon pairs. 
This unique environment provides exceptional sensitivity for tests of fundamental symmetries and for searches for rare BSM phenomena~\cite{Li:2016tlt}.

A major focus of recent experimental activity is the search for charge--parity (CP) violation through measurements of electric dipole moments (EDMs). 
The observation of a nonzero EDM would imply violation of both  Parity (P)
and Time-reversal (T) symmetries and would offer critical insight into the mechanism of baryogenesis. 
A key achievement highlighted in this review is the precision determination of the $\Lambda$ hyperon EDM using the entangled $\Lambda\bar{\Lambda}$ system~\cite{BESIII:2025vxm}. 
By exploiting spin correlations in $J/\psi \to \Lambda\bar{\Lambda}$ decays, this measurement reached a sensitivity at the level of $10^{-19}\,e\cdot\mathrm{cm}$, improving upon limits from experiments in the 1980s by three orders of magnitude~\cite{Pondrom:1981gu}. 
Complementary to CP-violation (CPV) studies, searches for baryon-number ($B$) and lepton-number ($L$) violation probe the fundamental structure of the SM. 
These include the first search for $\Lambda$--$\bar{\Lambda}$ oscillations~\cite{BESIII:2023tge, BESIII:2024gcd}, targeting $|\Delta B|=2$ transitions associated with Majorana dynamics, as well as systematic investigations of rare processes with $|\Delta B|=1$ and $|\Delta L|=2$, such as $\Sigma^- \to p e^- e^-$~\cite{BESIII:2020iwk} and multiple decay channels in the $\Xi$ and $\Lambda$ sectors~\cite{BESIII:2023str, BESIII:2025ylz}. 
Together, these measurements impose stringent constraints on BSM scenarios that predict violation of fundamental quantum numbers.

Hyperons also provide a sensitive window into possible dark-sector physics. 
If dark matter consists of light, weakly interacting states, it could appear experimentally as missing energy in rare hyperon decays. 
This possibility is further motivated by the neutron lifetime puzzle, in which discrepancies between experimental techniques may hint at baryon decays into invisible final states with branching fractions at the percent level~\cite{Fornal:2018eol}. 
BESIII has carried out pioneering searches along these lines, including the first exploration of dark-baryon production in $\Xi^- \to \pi^- + \text{invisible}$~\cite{BESIII:2025sfl} and searches for massless invisible particles in $\Sigma^+ \to p + \text{invisible}$~\cite{BESIII:2023utd}. 
In combination with new limits on invisible decays of the $\Lambda$ baryon~\cite{BESIII:2021slv}, these results establish world-leading constraints on models in which dark matter carries baryonic quantum numbers.

This review presents a comprehensive overview of the present experimental status of hyperon physics, emphasizing the implications of precision measurements for constraining BSM parameter space. 
We also discuss the future outlook, particularly the prospects offered by next-generation facilities such as the Super Tau--Charm Factory (STCF)~\cite{Achasov:2023gey}, which are expected to extend the sensitivity of hyperon-based probes to significantly deeper regimes in the ongoing search for the fundamental laws governing the universe.

\section{Precision Tests of Symmetries}

\subsection{Prob $\Lambda$ EDM Through the Entangled Strange Baryon-Antibaryon System}

The question of why the universe is dominated by matter rather than antimatter has driven fundamental research for decades. To generate this observed baryon asymmetry from a symmetric initial state, the Sakharov conditions necessitate the violation of CP symmetry and violation of baryon number conservation. However, the magnitude of CPV in the SM provided by the CKM mechanism, combined with the limits on the QCD vacuum angle $\bar{\theta}$, is insufficient to account for the magnitude of the matter-antimatter asymmetry observed in the universe~\cite{Sakharov:1967dj,Riotto:1998bt}. This discrepancy strongly suggests the existence of new physics (NP) beyond the SM that violates CP symmetry.
In the quest to uncover these new sources of CPV, the measurement of permanent EDMs of elementary particles has emerged as a sensitive and robust probe. A non-zero EDM for a spin-1/2 particle signifies a violation of both P and T symmetries. Assuming the conservation of CPT symmetry, T violation necessitates CP violation. Unlike CPV in flavor-changing processes, EDMs probe flavor-diagonal CPV, which is highly suppressed in the SM, making any observation of a significant EDM an unambiguous signal of new physics~\cite{Beacham:2019nyx,Chupp:2017rkp}.

Historically, experimental efforts have focused heavily on the neutron and atoms such as $^{199}\text{Hg}$, setting stringent upper limits on the QCD vacuum phase angle, $\bar{\theta}$~\cite{Abel:2020pzs, Graner:2016ses}. However, solely measuring the EDM of first-generation particles such as electrons and neutrons is insufficient to disentangle the potential sources of CPV, such as the quark EDM, the chromo-electric dipole moment (cEDM), and the $\bar{\theta}$ term~\cite{Pospelov:2005pr, Chupp:2017rkp, Chen:2025rab}. A global analysis requires inputs from diverse systems, including the heavy quark or strange quark sectors. Hyperons, which contain one or more strange valence quarks, offer a unique laboratory for this purpose. The strange quark may exhibit specific couplings to BSM fields, potentially enhancing the EDM effect in ways not accessible through neutron measurements~\cite{Chen:2025rab}.
Despite their theoretical importance, hyperon EDM measurements have been hindered by experimental challenges. Direct measurements utilizing spin precession in magnetic fields—standard for long-lived neutrons—are difficult for hyperons due to their extremely short lifetimes ($\sim 10^{-10}$ s). The only prior experimental limit for the $\Lambda$ hyperon, $|d_\Lambda| < 1.5 \times 10^{-16}$ e cm at the $95\%$ confidence level(CL), was established in a fixed-target experiment at Fermilab more than four decades ago~\cite{Pondrom:1981gu}.
Recent developments at electron-positron colliders, particularly the BESIII experiment, have opened a new avenue for precision hyperon physics~\cite{BESIII:2018cnd, BESIII:2021ypr, BESIII:2022qax, BESIII:2023drj, BESIII:2024nif}. By exploiting the quantum entanglement of hyperon-antihyperon pairs produced in the decay of the $J/\psi$ resonance, it is possible to extract EDM indirectly from the angular distributions of the decay products~\cite{He:1992ng, He:1993ar, He:2022jjc, Fu:2023ose, Du:2024jfc}.

In the context of effective field theory, contributions to the hyperon EDM can be parameterized by a Lagrangian~\cite{Chupp:2017rkp}:
\begin{eqnarray}
    \mathcal{L} = -\frac{\alpha_s}{8\pi}\bar{\theta} \text{Tr}[G^{\mu\nu}\tilde{G}_{\mu\nu}] - \frac{i}{2}\sum_q d_q \bar{q} F_{\mu\nu}\sigma^{\mu\nu}\gamma^5 q \nonumber \\
    - \frac{i}{2}\sum_q \tilde{d}_q \bar{q} G_{\mu\nu}^a T^a \sigma^{\mu\nu}\gamma^5 q.
\end{eqnarray}
Here, $d_q$ represents the quark EDM and $\tilde{d}_q$ represents the quark chromo-EDM. The $\Lambda$ hyperon EDM is sensitive to these parameters, particularly the strange quark contribution ($d_s$), which is largely unconstrained by neutron EDM measurements due to the dominance of $u$ and $d$ quarks in the nucleon wavefunction.  In electron-positron annihilation, hyperons are produced in pairs via the process $e^+e^- \rightarrow J/\psi \rightarrow B\bar{B}$. The vertex describing the interaction between the vector meson $J/\psi$ and the baryon pair can be parameterized by Lorentz-invariant form factors. The helicity amplitude for this decay is expressed as~\cite{He:2022jjc}:
\begin{eqnarray}
    \mathcal{M}_{\lambda_1, \lambda_2} = \epsilon_\mu(\lambda_1 - \lambda_2) \bar{u}(\lambda_1, p_1) ( F_V \gamma^\mu + \frac{i}{2m} \sigma^{\mu\nu} q_\nu H_\sigma \nonumber \\
    + \gamma^\mu \gamma^5 F_A + \sigma^{\mu\nu} \gamma^5 q_\nu H_T) v(\lambda_2, p_2).
\end{eqnarray}
In this decomposition, $F_V$ and $H_\sigma$ are the parity-conserving vector and tensor form factors, while $F_A$ and $H_T$ describe parity-violating and CP-violating effects, respectively. The term $H_T$ is of paramount importance for EDM searches, representing a CP-violating vertex in the $J/\psi B\bar{B}$ interaction. The CP-violating electric-dipole form factor is denoted by $d_B(q^2)$, and the static, or permanent, EDM is defined as its value at zero momentum transfer, $d_B\equiv d_B(0)$. In $J/\psi\to B\bar{B}$, the corresponding CP-violating form factor is probed at the timelike momentum transfer $q^2=M_{J/\psi}^2$ and can be written as
\begin{eqnarray}
    H_T(q^2) = \frac{2e}{3M_{J/\psi}^2} g_V d_B(q^2),
\end{eqnarray}
where $g_V$ is the coupling constant derived from the branching fraction of $J/\psi$ decay to lepton pairs. A non-zero measurement of $H_T$ serves as a direct indicator of CP violation in the production process. 

The BESIII Collaboration has reported a milestone measurement of the EDM of the $\Lambda$ hyperon, utilizing a sample of $10 \times 10^9$ $J/\psi$ events~\cite{BESIII:2025vxm}. This study circumvents traditional spin-precession challenges by employing the novel method mentioned before based on the spin-entangled $\Lambda\bar{\Lambda}$ system produced in $J/\psi \to \Lambda\bar{\Lambda}$ decays.
The analysis simultaneously extracted the CP-violating form factor $H_T$, the parity-violating form factor $F_A$, and the decay parameters. The result for the $\Lambda$ EDM was consistent with zero:
\begin{eqnarray}
\text{Re}(d_\Lambda) &=& (-3.1 \pm 3.2 \pm 0.5) \times 10^{-19} \, e \, \text{cm}, \nonumber \\
\text{Im}(d_\Lambda) &=& (2.9 \pm 2.6 \pm 0.6) \times 10^{-19} \, e \, \text{cm}.
\end{eqnarray}
This corresponds to an upper limit of $|d_\Lambda| < 6.5 \times 10^{-19} \, e \, \text{cm}$ at the $95\%$ CL, improving upon the previous world best limit by three orders of magnitude.  

\begin{figure}
    \centering
    \includegraphics[width=0.45\textwidth]{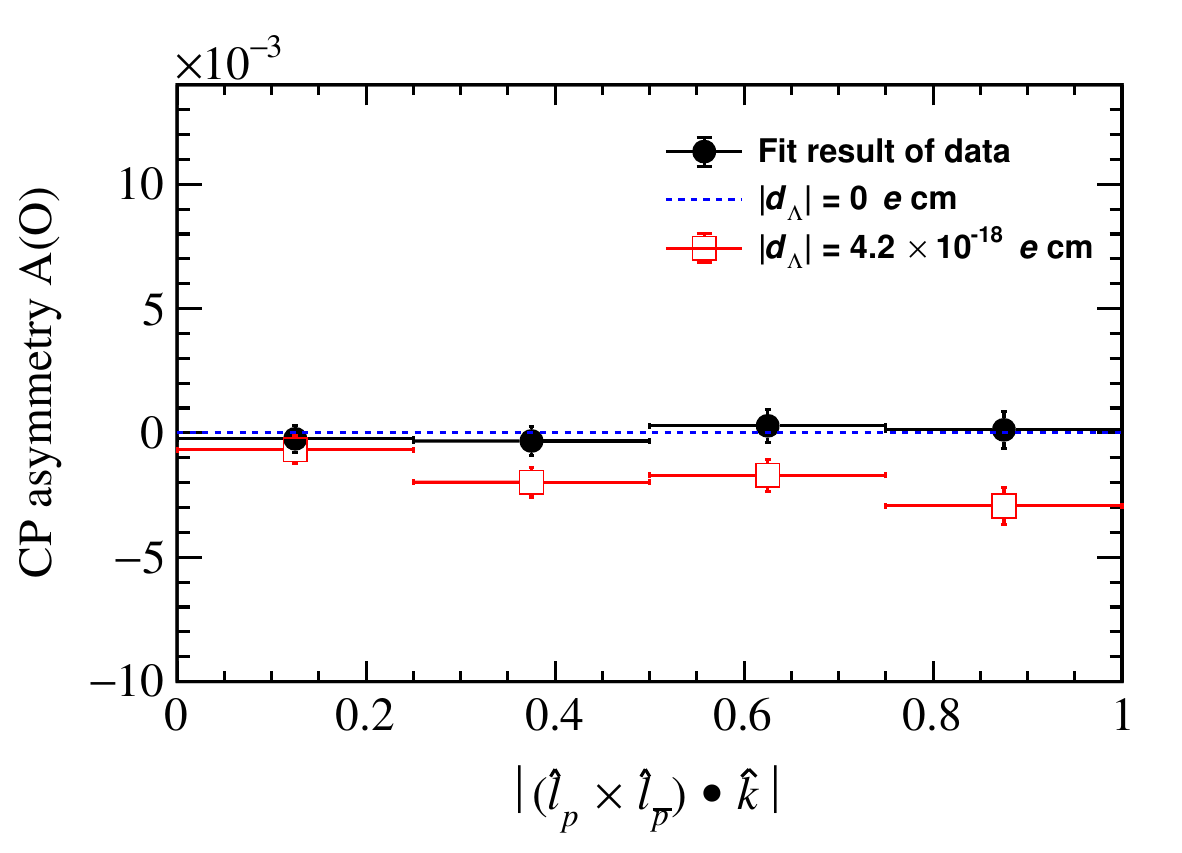}
    \caption{\textbf{EDM effects visualization}. The nominal fit, consistent with a null result within uncertainties, is shown by the black markers. In contrast, the red markers illustrate the expected distribution for a significant EDM of $|d_{\Lambda}| = 4.2 \times 10^{-18}\ e\text{ cm}$, manifesting as a $10\sigma$ departure from zero. The blue dashed line provides the theoretical baseline for a vanishing EDM.}
    \label{fig:todd}
\end{figure}

The experimental manifestation of the EDM effect in entangled $\Lambda\overline{\Lambda}$ pairs is observed through differences in their angular distributions. This can be quantified using the triple-product observable $O \equiv (\hat{l}_{p} \times \hat{l}_{\overline{p}}) \cdot \hat{k}$, where $\hat{l}_{p}$ ($\hat{l}_{\overline{p}}$) denotes the unit momentum of the proton (antiproton) in the $\Lambda$ ($\overline{\Lambda}$) rest frame, and $\hat{k}$ is the unit momentum of the $\Lambda$ in the $e^{+}e^{-}$ center-of-mass frame.
To visualize the experimental effect to the hyperon EDM, the analysis employs the CP asymmetry $A(O)$ derived from the triple product observable $O$, defined as:

\begin{equation}
A(O) = \frac{N_{\text{event}}(O>0)-N_{\text{event}}(O<0)}{N_{\text{event}}(O>0)+N_{\text{event}}(O<0)},
\end{equation}

where $N_{\text{event}}$ represents the number of events observed for each sign of $O$.  This observable is proportional to the real part of the CP-violating form factor $H_T$ and is shown in Fig.~\ref{fig:todd}.
The variation of $A(O)$ with respect to $|O|$ is evaluated using simulated events resampled from the nominal model based on the experimental fit. To contrast the current results with potential physics beyond the SM, a scenario is considered where the CP-violating form factor $H_T$ deviates by ten standard deviations from zero, corresponding to $|d_{\Lambda}| = 4.2 \times 10^{-18} \ e$ cm. This visualization highlights the distinction between the current measurement, which is consistent with zero, and a hypothetical non-zero EDM signal.

To interpret these findings at the energy scales of the $J/\psi$ resonance, a perturbative QCD analysis relates the $\Lambda$ EDM form factor $d_{\Lambda}(q^2)$ to fundamental CP-violating quark dipole interactions~\cite{Chen:2025rab}. This framework employs a collinear factorization formula where the form factor is expressed as a convolution of hard scattering coefficients with the light-cone distribution amplitudes (LCDAs) of the hyperon. The analysis indicates that $d_{\Lambda}(q^2)$ follows a $(q^2)^{-4}$ scaling behavior in the high-energy limit, which is consistent with general power-counting rules for exclusive processes.

\begin{figure*}[t!!]
  \centering
  \subfigure[]{
  {\label{fig:bound1}}
  \includegraphics[width=0.45\linewidth]{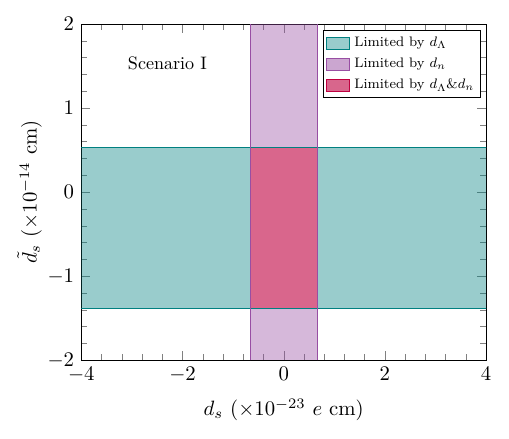}}
  \subfigure[]{
  {\label{fig:bound2}}
  \includegraphics[width=0.45\linewidth]{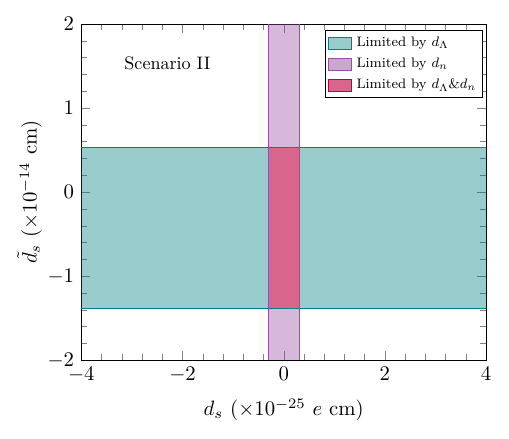}}
  \caption{Constraints on the $s$-quark EDM $d_s$ and CEDM $\tilde d_s$ from the measurements on the $\Lambda$ and $n$ EDMs. (a) Scenario I: $d_s \gg d_u, d_d$ ; (b) Scenario II: $d_s=d_u=d_d$.
  }
  \label{fig:bound}
\end{figure*}

The physical significance of the $\Lambda$ EDM lies in its unique sensitivity to the strange-quark sector, providing a probe that is complementary to the neutron EDM. Figure~\ref{fig:bound} displays the constraints on the $d_s$ and CEDM $\tilde{d}_s$ derived from current experimental limits.
The $\Lambda$ EDM bound is specifically sensitive to the $\tilde{d}_s$, whereas the neutron EDM provides a more stringent constraint on the $d_s$. While the $\Lambda$ EDM constraint alone results in a linear correlation between $d_s$ and $\tilde{d}_s$ appearing as an elongated band, the integration of neutron EDM data significantly reduces this parameter space. Because the neutron EDM lacks dependence on the $\tilde{d}_s$, its tight experimental limit restricts $d_s$, allowing the hyperon data to isolate the effect of $\tilde{d}_s$. Under these combined constraints, a new upper bound on the $\tilde{d}_s$ is established at
\begin{equation}
    |\tilde{d}_s| \le 1.4 \times 10^{-14}\rm\ cm.
\end{equation}
These results highlight the essential role of hyperon measurements in a global search for flavor-diagonal CP violation beyond the SM.

\subsection{Search for Baryon and Lepton Number Violating Process}
Although the SM preserves baryon and lepton number as accidental global symmetries at the perturbative level, non-perturbative effects and various grand unified theories (GUTs) suggest that these conservation laws are merely consequences of low-energy effective operators\cite{Weinberg:1979sa}.

Minimal GUT frameworks, such as $SU(5)$, predominantly feature $B-L$ conserving interactions at the dimension-six level, which lead to proton decay signatures~\cite{Georgi:1974sy}. However, the absence of conclusive experimental evidence for such processes~\cite{Super-Kamiokande:2009yit} motivates the exploration of alternative symmetry-breaking patterns, including $(B-L)$-violating transitions arising from dimension-seven operators or $|\Delta(B-L)|=2$ interactions~\cite{Babu:2012iv}. These searches are deeply intertwined with the investigation of neutrino nature; if neutrinos are Majorana fermions~\cite{Gribov:1968kq}, lepton number violation process provides a natural signature of the seesaw mechanism~\cite{Mohapatra:1979ia} and offers a unique channel to study neutrino mass generation through hyperon-based analogs of neutrinoless double beta decay~\cite{Dolinski:2019nrj}.

Hyperon systems, incorporating second-generation strange quarks, provide an essential laboratory for probing the flavor structures of baryon and lepton number non-conservation~\cite{Kang:2009xt, Li:2016tlt}. Unlike nucleon decay, hyperon transitions allow for the investigation of complex interference patterns among multiple baryon number violation (BNV) or lepton number violation (LNV) amplitudes, which are sensitive to the relative phases of different operators~\cite{McCracken:2015coa}. Furthermore, processes such as $\Delta B=2$ hyperon-antihyperon oscillations and $|\Delta L|=2$ rare decays offer stringent tests of physics beyond the SM, providing experimental constraints that are critical for refining theoretical models of baryogenesis~\cite{Hernandez-Tome:2021byt, Hernandez-Tome:2022ejd}.

The BESIII Collaboration initiated a systematic search for forbidden processes in the hyperon sector, beginning with the investigation of $\Sigma^-$ decays using a sample of $1.31 \times 10^9$ $J/\psi$ events~\cite{BESIII:2020iwk}. This study focused on the $|\Delta L|=2$ LNV decay $\Sigma^- \to p e^- e^-$ and the rare inclusive decay $\Sigma^- \to \Sigma^+ X$, where $X$ represents invisible or undetected particles. The search for $\Sigma^- \to p e^- e^-$ is particularly significant as it tests the $\Delta Q = \Delta L = 2$ rule, a process that could be mediated by the exchange of Majorana neutrinos~\cite{Littenberg:1991rd}.
The experimental analysis utilized the process $J/\psi \to \Sigma^- \bar{\Sigma}^+$, where the $\bar{\Sigma}^+$ anti-hyperons were reconstructed through their dominant $\bar{p} \pi^0$ and $n \pi^-$ decay modes to provide a clean environment for searching for the signal on the recoiling side. To ensure high background suppression, a kinematic fit was applied, and the signal yields were extracted from the distribution of the recoil mass of the tagged $\bar{\Sigma}^+$. No significant signal was observed in either channel, significantly improving upon previous limits set by the HyperCP experiment~\cite{HyperCP:2005sby}. The upper limits on the branching fractions at the 90\% CL were determined to be:
\begin{equation}
\begin{aligned}
    \mathcal{B}(\Sigma^- \to p e^- e^-) &< 6.7 \times 10^{-5},  \\ \mathcal{B}(\Sigma^- \to \Sigma^+ X) &< 1.2 \times 10^{-4}
\end{aligned}
\end{equation}
These results represent the first experimental constraints on these specific $\Sigma^-$ decay modes, and the results are well above the theoretical predictions~\cite{Barbero:2002wm, Barbero:2007zm, Barbero:2013fc,Zhao:2026CPL}. Recent covariant baryon chiral perturbation theory calculations have provided a systematic effective-field-theory treatment of spin-$1/2$ hyperon neutrinoless double-beta decays, including all accessible channels and the dependence on the effective Majorana neutrino mass~\cite{Zhao:2026CPL}.

Utilizing a massive dataset of $10 \times 10^9$ $J/\psi$ events, the BESIII Collaboration expanded the search to the double-strange $\Xi^0$ hyperon system~\cite{BESIII:2023str}. This analysis targeted the BNV and LNV decays $\Xi^0 \to K^- e^+$ and $\Xi^0 \to K^+ e^-$, which explore different selection rules regarding the $B-L$ symmetry: the former satisfies $\Delta(B-L)=0$ while the latter involves $|\Delta(B-L)|=2$.  The study employed a double-tag method, where the $\bar{\Xi}^0$ was reconstructed through $\bar{\Xi}^0 \to \bar{\Lambda}(\to \bar{p}\pi^+) \pi^0(\to \gamma\gamma)$. The signal search was performed by analyzing the invariant mass of the $K^\pm e^\mp$ candidates and the energy difference $\Delta E = E_{\rm sig} - E_{\rm beam}$. With no signal candidates found, the collaboration established the following upper limits at the 90\% CL:
\begin{equation}
\begin{aligned}
    \mathcal{B}(\Xi^0 \to K^- e^+) &< 3.6 \times 10^{-6}, \\
    \mathcal{B}(\Xi^0 \to K^+ e^-) &< 1.9 \times 10^{-6}
\end{aligned}
\end{equation}
These values constitute the first experimental limits on these decays, and offer a direct probe of the BNV
processes which strange quarks involved.

In 2023, the BESIII Collaboration reported a search for $\Lambda-\bar{\Lambda}$ oscillations, a $|\Delta B|=2$ process that provides a unique test for baryon number violation involving second-generation quarks~\cite{BESIII:2023tge}. The time evolution of such a system is described by a mass matrix $M$ where the off-diagonal elements $\delta m_{\Lambda\bar{\Lambda}}$ represent the BNV transition:
\begin{equation}
    M = \begin{pmatrix} m_{\Lambda} - \Delta E_{\Lambda} & \delta m_{\Lambda\bar{\Lambda}} \\ \delta m_{\Lambda\bar{\Lambda}} & m_{\bar{\Lambda}} - \Delta E_{\bar{\Lambda}} \end{pmatrix}
\end{equation}
The experiment utilized the coherent production of $\Lambda\bar{\Lambda}$ pairs from $J/\psi$ decays, specifically $J/\psi \to p K^- \bar{\Lambda} + c.c.$, which allows for the detection of "wrong-sign" events ($J/\psi \to p K^- \Lambda$) that would signify an oscillation~\cite{Kang:2009xt}.
Based on $1.31 \times 10^9$ $J/\psi$ events, the analysis found zero wrong-sign events in the signal region. The upper limit for the oscillation probability $\mathcal{P}(\Lambda)$ was set at $4.4 \times 10^{-6}$ at the 90\% C.L. This result was used to derive a constraint on the oscillation parameter $\delta m_{\Lambda\bar{\Lambda}}$ using the relation $(\delta m_{\Lambda\bar{\Lambda}})^2 = \mathcal{P}(\Lambda) / (2\tau_\Lambda^2)$, where $\tau_\Lambda$ is the $\Lambda$ lifetime. The findings yielded:
\begin{equation}
\begin{aligned}
    \delta m_{\Lambda\bar{\Lambda}} &< 3.8 \times 10^{-18} \text{ GeV}, \\ \tau_{osc} &> 1.7 \times 10^{-7} \text{ s}
\end{aligned}
\end{equation}
This measurement provides a critical benchmark for $|\Delta B|=2$ interactions and serves as a vital comparison to neutron-antineutron oscillation experiments.
More recently, BESIII has tightened these constraints by analyzing a significantly larger dataset of $10 \times 10^{9}$ $J/\psi$ events~\cite{BESIII:2024gcd}. Employing the same coherent production method in the decay channel $J/\psi \to \Lambda \bar{\Lambda}$, the updated analysis establishes a stronger upper bound on the integrated oscillation probability, $\mathcal{P}(\Lambda) < 1.4 \times 10^{-6}$, and on the transition mass, $\delta m_{\Lambda\bar{\Lambda}} < 2.1 \times 10^{-18},\mathrm{GeV}$, both quoted at the $90\%$ CL. These results imply a lower limit on the oscillation time of $\tau_{\mathrm{osc}} > 3.1 \times 10^{-7}\ \mathrm{s}$.
These findings provide a new experimental benchmark for testing BNV theories, with the potential for significantly improved sensitivity at future next-generation super $\tau$-charm factories.

The most recent investigation in this series is the first search for the LNV process $\Xi^- \to \Sigma^+ e^- e^-$~\cite{BESIII:2025ylz}. This decay, characterized by $\Delta L = 2$ and $\Delta S = 1$, is a hyperon analog of neutrinoless double-beta decay and is sensitive to the exchange of Majorana neutrinos~\cite{Hernandez-Tome:2022ejd}. The analysis utilized the full $10 \times 10^9$ $J/\psi$ dataset, employing a double-tag technique through $J/\psi \to \Xi^- \bar{\Xi}^+$ with $\bar{\Xi}^+ \to \bar{\Lambda} \pi^+$ and $\bar{\Lambda} \to \bar{p} \pi^+$.
A blind analysis strategy was adopted to avoid human bias. After applying event selection and background suppression using the recoil mass against the $\bar{\Xi}^+$ and the invariant mass of the $\Sigma^+ e^- e^-$ system, no significant excess of signal events was found. The upper limit on the branching fraction was established at the 90\% CL as:
\begin{equation}
    \mathcal{B}(\Xi^- \to \Sigma^+ e^- e^-) < 2.0 \times 10^{-5}
\end{equation}
This result represents the first experimental constraint on this channel and provides an important benchmark for related theoretical studies. In particular, the recent chiral effective-field-theory calculation predicts branching fractions many orders of magnitude below current experimental sensitivities for the light-Majorana-neutrino long-range contribution, while emphasizing the role of short-range counterterms in a complete description~\cite{Zhao:2026CPL}.

\section{Experimental Searches for the Dark Sector}

The search for the dark sector is a fundamental frontier in modern particle physics, driven by the persistent mysteries of dark matter (DM) and the observed matter-antimatter asymmetry of the Universe~\cite{Sakharov:1967dj}. While astronomical observations provide robust indirect evidence for DM, direct detection in terrestrial experiments remains elusive~\cite{Planck:2018vyg}. A compelling theoretical hint lies in the striking similarity between the dark matter and baryon mass densities:
$\rho_{DM} \approx 5.4\rho_{baryon}$~\cite{Planck:2018vyg}.
This coincidence suggests a common origin for both sectors, potentially linked through processes that violate baryon number conservation~\cite{Petraki:2013wwa}.

In the asymmetric dark matter scenario, DM particles with masses at the GeV scale could be related to the baryon asymmetry, potentially interacting with the SM through a neutron portal operator $udd$~\cite{Zurek:2013wia}. This framework has gained significant attention as a possible resolution to the "neutron lifetime puzzle"—the $1\%$ discrepancy between neutron lifetime measurements using the beam and bottle methods. Such a discrepancy could be explained if neutrons possess a branching fraction of approximately $1\%$ into dark sector particles~\cite{Petraki:2013wwa, Fornal:2018eol}.

Beyond the neutron, the hyperon sector provides a unique laboratory to test these theories~\cite{Goudzovski:2022vbt}. Rare flavor-changing neutral current (FCNC) transitions, such as the $s \rightarrow d$ quark transition, are extremely suppressed in the SM with predicted branching fractions below $10^{-11}$~\cite{Buchalla:1995vs, Tandean:2019tkm}. However, new physics models involving massless dark photons ($\gamma'$), QCD axions ($a$), or dark baryons ($\chi$) can enhance these BFs to the order of $10^{-4}$, providing a clear signature of missing energy~\cite{Su:2019ipw,MartinCamalich:2020CPL,MartinCamalich:2025CPL}. In particular, flavor-violating QCD axion scenarios make baryon and hyperon decays a sensitive and complementary probe of $s\to d$ axion couplings~\cite{MartinCamalich:2020CPL}.

Recently NA62 has reported~\cite{NA62:2018ctf}  $\mathcal{B}(K^{+} \rightarrow \pi^{+} \nu \bar{\nu}) = (13.0^{+3.3}_{-3.0}) \times 10^{-11}$. This result is consisitent with SM prediction, but allow the final invisible states to be dark matter pairs~\cite{He:2025sao}, which further motivate the search for similar signatures in hyperon decays. Unlike hadron colliders where missing energy is difficult to constrain due to high backgrounds and unknown initial states~\cite{Borsato:2021aum}, $e^+ e^-$ experiments like BESIII offer a clean environment with well-defined production processes, such as $J/\psi \rightarrow B\bar{B}$, where the reconstruction of one baryon allows for a nearly model-independent search for the invisible decay of its partner.

\begin{figure}[h]
    \centering
    \includegraphics[width=0.45\textwidth]{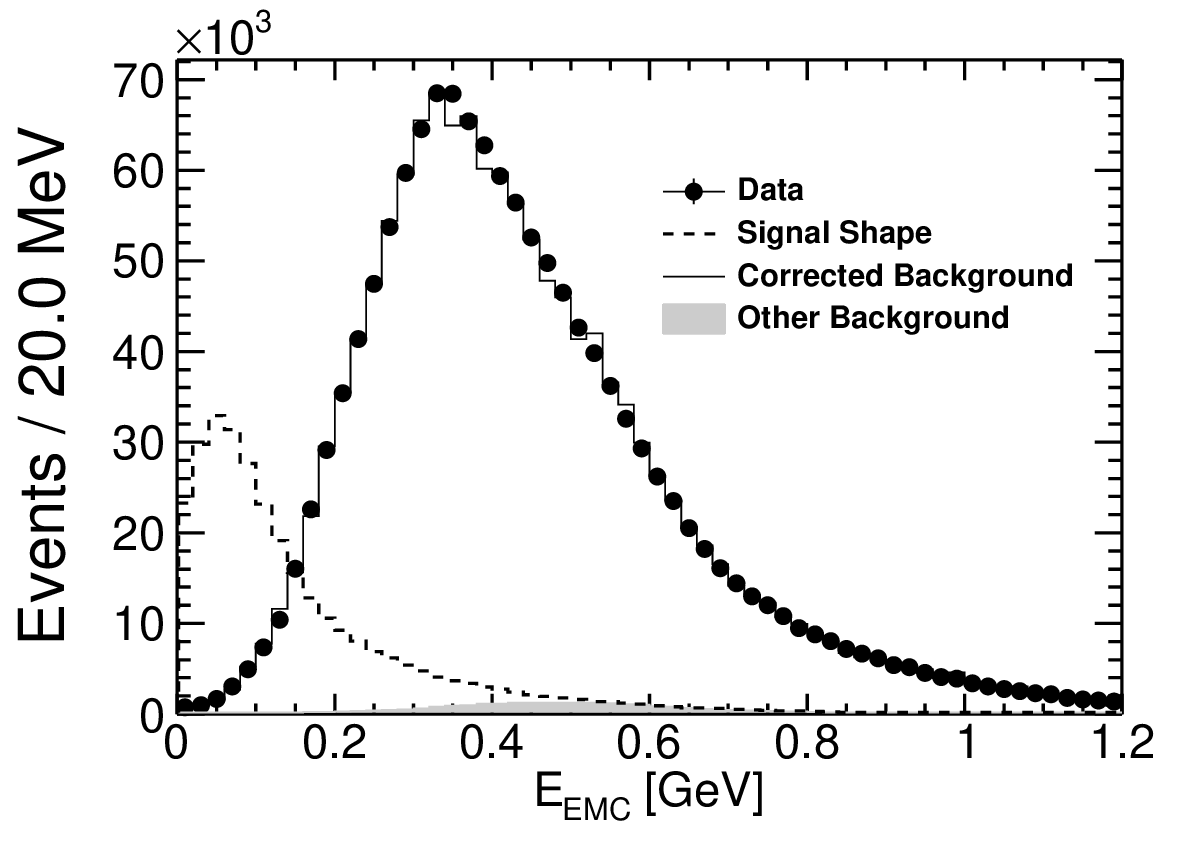}
    \caption{The $E_{\text{EMC}}$ distribution for the $\Lambda \rightarrow \text{invisible}$ search. The dots with uncertainties represent data, while the solid line shows the $\Lambda \rightarrow n \pi^0$ background shape after data-driven corrections. The dashed line indicates the expected signal shape (normalized arbitrarily for clarity), which peaks near zero energy deposition.}
    \label{fig:lambda_eemc}
\end{figure}

\subsection{First Search for Invisible Baryon Decays: $\Lambda \to \text{invisible}$}

The search for the dark sector through baryonic transitions began with a pioneering study of the invisible decays of the $\Lambda$ hyperon, utilizing $10 \times 10^{9}$ $J/\psi$ events collected with the BESIII detector~\cite{BESIII:2021slv}. The experiment employs a model-independent double-tag strategy in the process $J/\psi \rightarrow \Lambda \bar{\Lambda}$. One $\bar{\Lambda}$ baryon is explicitly reconstructed through its dominant hadronic decay $\bar{\Lambda} \rightarrow \bar{p} \pi^+$, effectively "tagging" the presence of a recoiling $\Lambda$ baryon. A binned maximum likelihood fit to the recoil mass of the $\bar{p}\pi^+$ system yields a total of $N_{\text{tag}} = 4,154,428 \pm 2,040$ single-tag events. The signal for $\Lambda \rightarrow \text{invisible}$ is then defined by the absence of any additional activity in the detector associated with the recoiling $\Lambda$.

The central discriminator for extracting the signal is $E_{\text{EMC}}$, defined as the sum of energies of all showers in the electromagnetic calorimeter (EMC) not associated with the tagged charged tracks. Ideally, for an invisible decay, this value should be near zero, comprised only of detector noise and minor leakage from charged particles. The distribution is modeled by the following components:
\begin{equation}
E_{\text{EMC}} = E_{\text{EMC}}^{\pi^0} + E_{\text{EMC}}^{n} + E_{\text{EMC}}^{\text{noise}} \text{,}
\end{equation}
where $E_{\text{EMC}}^{\pi^0}$ accounts for electromagnetic showers from background $\pi^0$ decays, and $E_{\text{EMC}}^{n}$ represents energy deposited by neutrons. The dominant background originates from the SM process $\Lambda \rightarrow n \pi^0$. Because the hadronic interactions of neutrons in the detector material are difficult to simulate accurately using GEANT4, BESIII utilized a data-driven neutron control sample from $J/\psi \rightarrow \bar{p} \pi^+ n \pi^0$ events to correct the simulated background shapes.

The branching fraction is determined by comparing the observed double-tag yield against the single-tag yield and the relative detection efficiency $\epsilon_{\text{sig}}/\epsilon_{\text{tag}}$. After performing a binned maximum likelihood fit to the $E_{\text{EMC}}$ distribution (as shown in Fig.~\ref{fig:lambda_eemc}), no significant signal was observed. Including all systematic and statistical uncertainties, the upper limit at the 90\% CL is found to be:
\begin{equation}
\mathcal{B}(\Lambda \rightarrow \text{invisible}) < 7.4 \times 10^{-5} \text{.}
\end{equation}
This result represents the first direct experimental constraint on invisible baryon decays. It is consistent with the prediction of $4.4 \times 10^{-7}$ from mirror matter models~\cite{Tan:2020gpd} and serves to sharpen our understanding of dark sector models related to baryon asymmetry.

\subsection{Search for Massless Dark Particles: $\Sigma^{+} \to p + \text{invisible}$}

Another study, published in 2024, utilized the same sample of $10 \times 10^{9}$ $J/\psi$ events to search for massless particles beyond the SM, such as massless dark photons ($\gamma^{\prime}$) or QCD axions ($a$) in the $\Sigma^{+} \to p + \text{invisible}$ process~\cite{BESIII:2023utd}. The QCD axion interpretation is motivated by flavor-violating axion frameworks in which hyperon decays provide a clean probe complementary to kaon and meson-mixing observables~\cite{MartinCamalich:2020CPL,MartinCamalich:2025CPL}.
This analysis utilized the double-tag technique within the $J/\psi \to \Sigma^{+} \bar{\Sigma}^{-}$ production process. The $\bar{\Sigma}^{-}$ candidates were first reconstructed through the dominant $\bar{\Sigma}^{-} \to \bar{p} \pi^{0}$ decay to form a single-tag (ST) sample. The signal $\Sigma^{+} \to p + \text{invisible}$ was then searched for in the system recoiling against the ST $\bar{\Sigma}^{-}$. The absolute BF of the signal decay is determined by:
$
B_{sig} = \frac{N_{DT}^{obs}/\epsilon_{DT}}{N_{ST}^{obs}/\epsilon_{ST}} \text{,}
$
where $N_{ST}^{obs} (N_{DT}^{obs})$ is the observed ST (double-tag, DT) yield and $\epsilon_{ST} (\epsilon_{DT})$ is the corresponding detection efficiency.

\begin{figure}[t!!]
  \centering
  \subfigure[]{
  {\label{fig:Sigma_Eextra_1}}
  \includegraphics[width=0.95\linewidth]{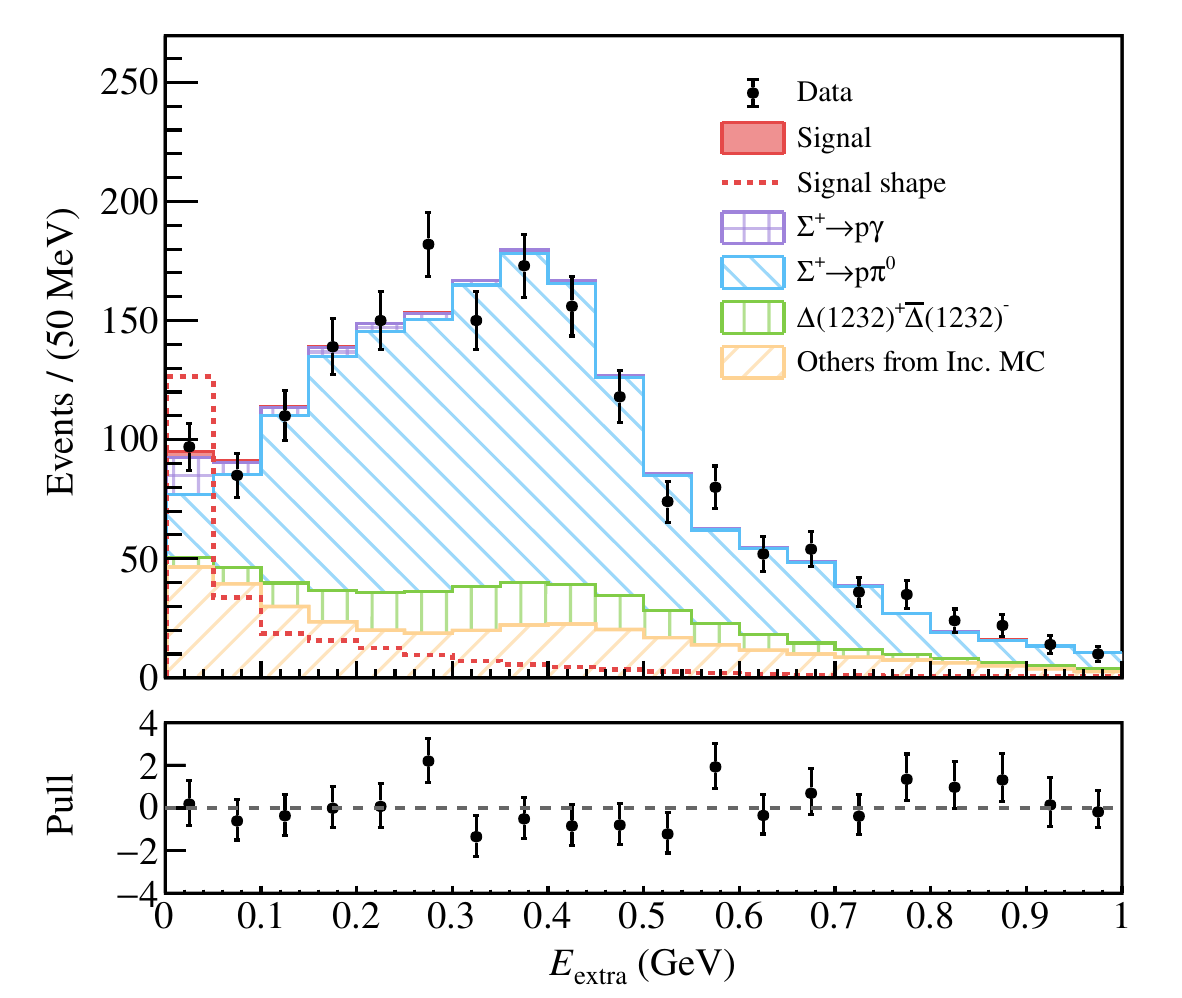}}

  \subfigure[]{
  {\label{fig:Sigma_Eextra_2}}
  \includegraphics[width=0.95\linewidth]{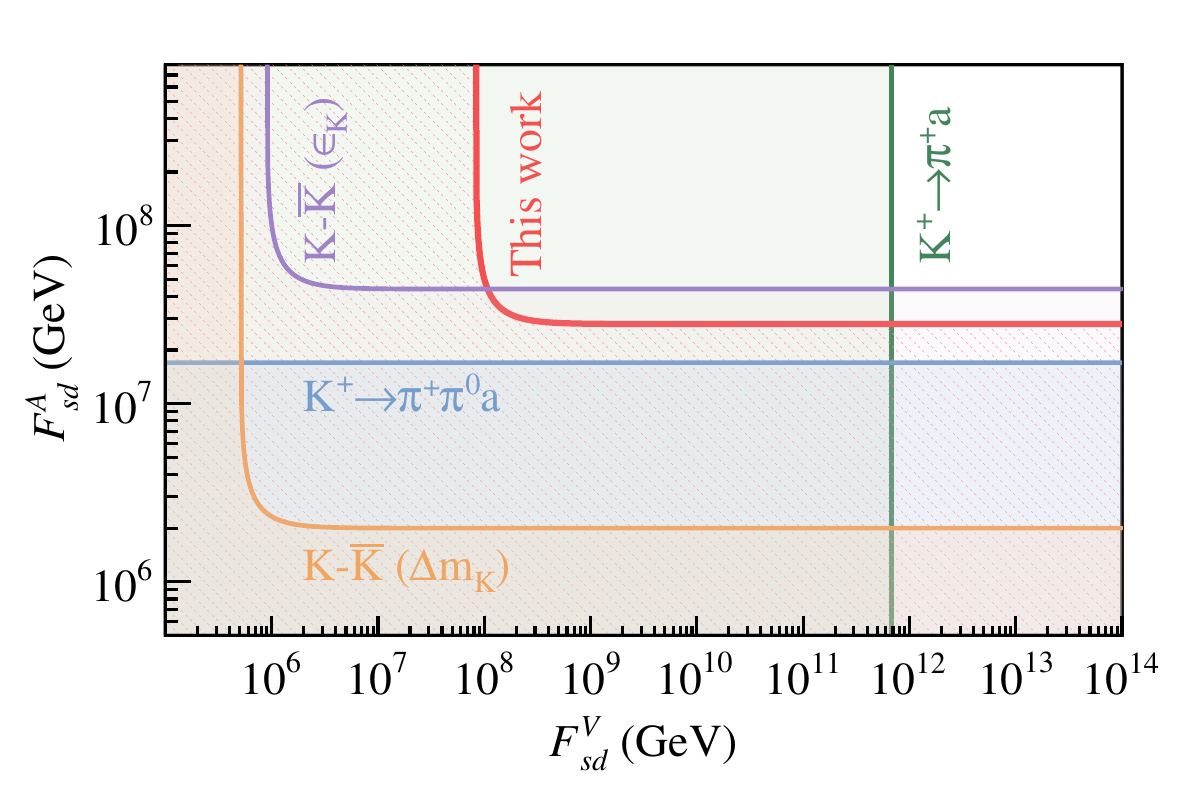}}
  \caption{(a) Fitted $E_{\text{extra}}$ spectra for the process $\Sigma^+ \rightarrow p + \text{invisible}$; (b) Constraints at the $90\%$ C.L. on the effective axion--fermion decay constant associated with the $s \to d$ transition, as determined in this analysis. The shaded hatched area represents the region of parameter space excluded by the measurement. The quantities $F^{V}_{sd}$ and $F^{A}_{sd}$ correspond to the vector and axial-vector contributions to the decay constant, respectively.
  }
  \label{fig:Sigma_Eextra}
\end{figure}

The primary discriminator for extracting the DT yield is , the sum of energy from all showers in the EMC except for those associated with the ST side. This energy is partitioned as:
\begin{equation}
E_{\rm extra} = E_{\rm extra}^{\rm DT\pi^0/\gamma} + E_{\rm extra}^{\rm other} \text{,}
\end{equation}
where  denotes the energy from potential background photons or  on the DT side, and  accounts for other sources such as noise or interaction leakage. As no significant signal was observed in the  distribution which shown in Fig.~\ref{fig:Sigma_Eextra_1}, an upper limit at the $90\%$ CL was set:
\begin{equation}
B(\Sigma^{+} \to p + \text{invisible}) < 3.2 \times 10^{-5} \text{.}
\end{equation}

This result imposes stringent limits on new physics models, as shown in Fig.~\ref{fig:Sigma_Eextra_2}. Specifically, it constrains the axial-vector part $F_{sd}^{A}$, defined as the effective decay constant associated with the axial-vector $s\to d$ axion--fermion coupling, to be $F_{sd}^{A}>2.8\times10^7$ GeV. As in Ref.~\cite{BESIII:2023utd}, the comparison includes constraints from $K^+\to\pi^+a$, $K^+\to\pi^+\pi^0a$, and $K-\bar{K}$ mixing~\cite{E787:2000iwe,ParticleDataGroup:2022pth,UTfit:2022hsi}. The treatment of the flavor-violating axion couplings, in particular the extraction of the $K-\bar{K}$ mixing constraints under the assumption that the unknown low-energy constants vanish, follows Ref.~\cite{MartinCamalich:2020CPL}. More recent discussions of light flavor-violating dark sectors and updated kaon-sector constraints can be found in Refs.~\cite{MartinCamalich:2025CPL,NA62:2025CPL}.

\subsection{Search for Dark Baryons and Testing Baryon Number Conservation: $\Xi^{-} \to \pi^{-} + \text{invisible}$}

While the earlier investigations in the $\Lambda$ and $\Sigma$ sectors primarily addressed massless dark sector candidates or standard invisible final states, the 2026 BESIII study extended this frontier to the search for a massive dark baryon ($\chi$)~\cite{BESIII:2025sfl}. Motivated by theoretical frameworks such as B-Mesogenesis~\cite{Nelson:2019CPL,Elor:2018twp} and the neutron lifetime puzzle~\cite{Fornal:2018eol}, which posit a dark sector carrying baryon number, this experiment utilized $10 \times 10^9$ $J/\psi$ events to probe the decay $\Xi^- \to \pi^- + \text{invisible}$. This process provides a sensitive laboratory to search for dark baryons with masses ($m_{\chi}$) near the GeV/$c^2$ scale.

The analysis employs a double-tag method in the production process $J/\psi \to \Xi^- \bar{\Xi}^+$. The single-tag $\bar{\Xi}^+$ candidates are reconstructed via the hadronic decay chain $\bar{\Xi}^+ \to \bar{\Lambda} \pi^+ \to \bar{p} \pi^+ \pi^+$. In the system recoiling against the tagged $\bar{\Xi}^+$, the signal is identified by the presence of a single charged pion track ($\pi^-$) and the absence of additional energy deposition. The primary discriminator for signal extraction is $E_{\text{EMC}}$, the sum of energies of all showers in the EMC not associated with the tagged particles. For a dark baryon signal, $E_{\text{EMC}}$ should peak near zero, whereas SM backgrounds such as $\Xi^- \to \Lambda \pi^-$ and $\Xi^- \to \Sigma^0 \pi^-$ deposit significant energy through neutral fragments or secondary interactions. 

This analysis uses a data-driven control sample to accurately model the background $E_{\text{EMC}}$ distribution. The dominant background arises from $\Xi^- \to \Lambda(\to n \pi^0) \pi^-$, where the neutral fragments ($n, \pi^0$) mimic the invisible signature if their energy deposition is low. To mitigate the discrepancies between GEANT4 simulation and real data regarding hadronic interactions, a control sample of $\Xi^- \to \Lambda(\to p \pi^-) \pi^-$ was selected. By comparing the $E_{\text{EMC}}$ response of this high-purity hadronic sample between data and Monte Carlo, the researchers derived correction factors for the signal-region $E_{\text{EMC}}$ shape. This application of data-driven PDF corrections significantly reduced systematic uncertainties related to detector noise and neutral particle response.

\begin{figure}[htbp!!]
  \centering
  \subfigure[]{
  {\label{fig:Xi_1}}
  \includegraphics[width=0.95\linewidth]{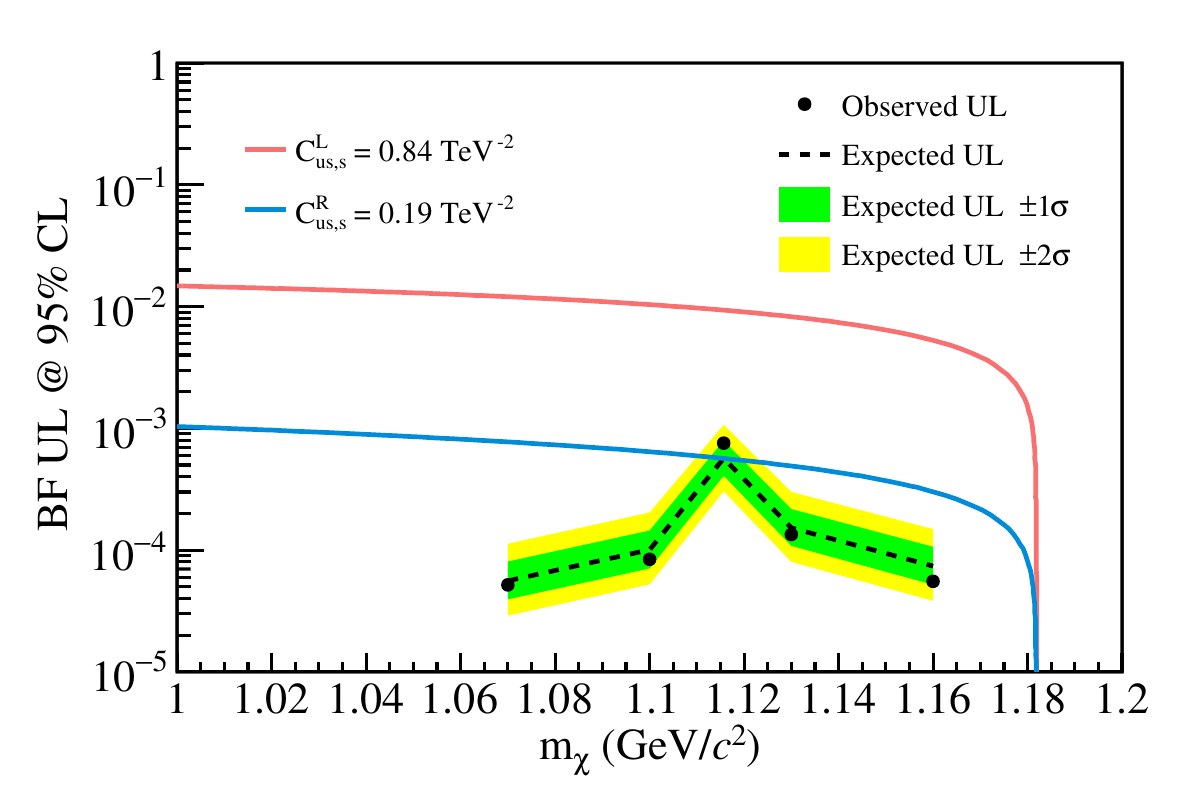}}

  \subfigure[]{
  {\label{fig:Xi_2}}
  \includegraphics[width=0.95\linewidth]{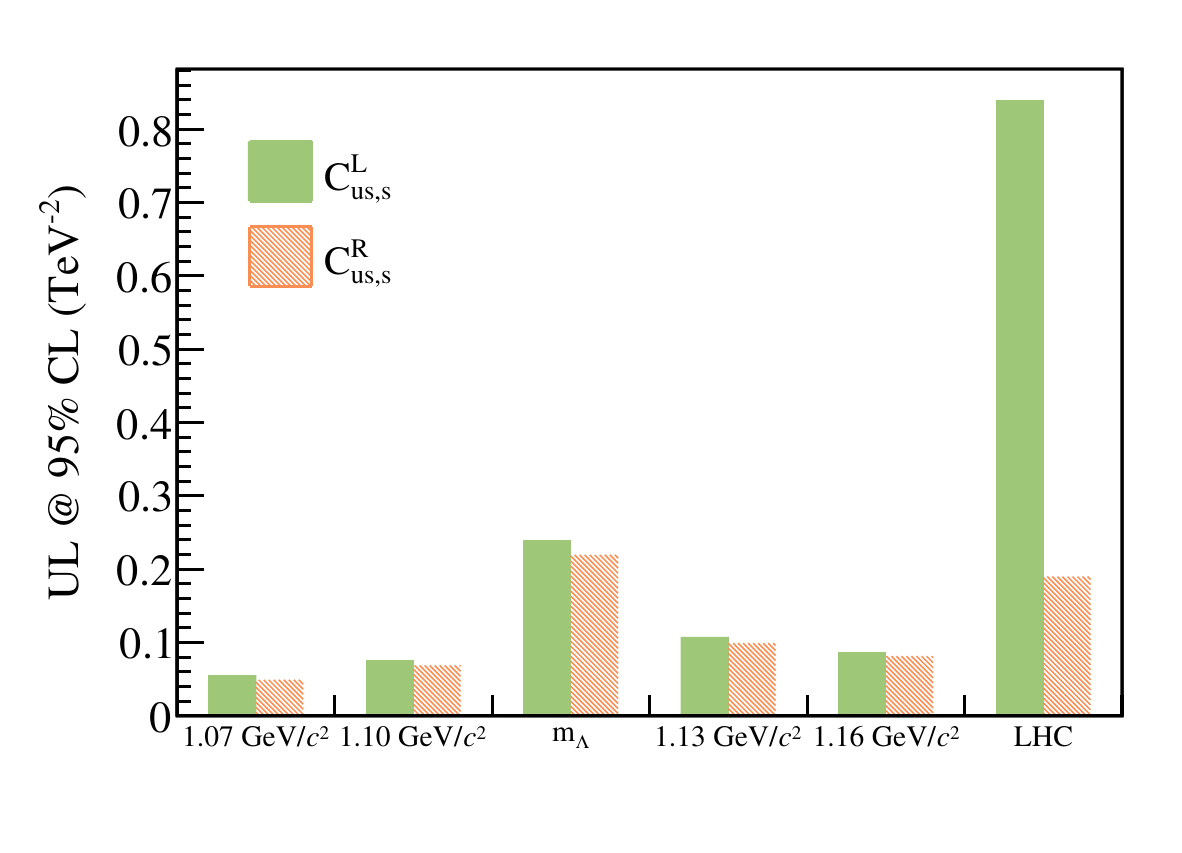}}
  \caption{Summary of 95\% C.L. constraints for $\Xi^- \to \pi^- + \text{invisible}$ 
    under different $m_\chi$ assumptions: (a) upper limits on the branching fraction, 
    and (b) the corresponding bounds on the Wilson coefficients 
    $C_{us,s}^{L}$ and $C_{us,s}^{R}$, with LHC results shown for comparison.
  }
  \label{fig:Xi}
\end{figure}

The search was systematically performed across five mass hypotheses: $m_\chi = 1.07, 1.10, m_\Lambda, 1.13$, and $1.16$ GeV/$c^2$.
No significant signal excess was observed in the $E_{\text{EMC}}$ distribution for any of the mass hypotheses. Consequently, upper limits on the branching fraction were established at the 90\% CL. For the hypothesis where the dark baryon mass equals the $\Lambda$ baryon mass ($m_\chi = m_\Lambda$), the limit is found to be:
\begin{equation}
\mathcal{B}(\Xi^- \to \pi^- + \text{invisible}) < 6.5 \times 10^{-4} \text{.}
\end{equation}
As shown in Fig.~\ref{fig:Xi_1}, the upper limits vary between $4.5 \times 10^{-5}$ and $1.1 \times 10^{-3}$ depending on the mass point. These results represent the first direct constraints on dark baryons from $\Xi^-$ decays and provide the most stringent limits to date on the Wilson coefficients of effective operators mediating quark-to-dark-baryon transitions in most cases, as shown in Fig.~\ref{fig:Xi_2}. For the right-handed coefficient $C^R_{us,s}$ at $m_\chi=m_\Lambda$, however, the BESIII constraint does not surpass the previous bound from LHC searches~\cite{Alonso-Alvarez:2021oaj}.

\begin{table*}[hbtp!]
\caption{Summary of experimental results for new physics searches in the hyperon sector at BESIII. The limits represent the most stringent constraints obtained for the corresponding channels.}
\begin{ruledtabular}
\begin{tabular}{llc}
Physics Topic & Observable & Experimental Limit (90\% C.L.) \\
\hline
Dark Baryon Sector & $\mathcal{B}(\Xi^- \to \pi^- \chi)$ & $4.2 \times 10^{-5} - 6.5 \times 10^{-4}$ \\
\hline
Invisible Decay & $\mathcal{B}(\Lambda \to \text{invisible})$ & $7.4 \times 10^{-5}$  \\
\hline
Massless BSM Particle & $\mathcal{B}(\Sigma^+ \to p + \text{invisible})$ & $3.2 \times 10^{-5}$ \\
\hline
$\Lambda$ EDM & $|d_\Lambda|$ & $6.5 \times 10^{-19} e \cdot \text{cm}$ (95\% C.L.) \\
\hline
Baryon/Lepton Violation & $\mathcal{B}(\Xi^0 \to K^- e^+)$ & $3.6 \times 10^{-6}$ \\
 & $\mathcal{B}(\Xi^- \to \Sigma^+ e^- e^-)$ & $2.0 \times 10^{-5}$ \\
 & $\mathcal{B}(\Sigma^- \to p e^- e^-)$ & $6.7 \times 10^{-5}$  \\
 & $\delta m_{\Lambda\bar{\Lambda}}$ & $3.8 \times 10^{-18}$ GeV  \\
\end{tabular}
\end{ruledtabular}
\label{tab:sum}
\end{table*}

\section{Summary and Conclusion}

A summary of experimental results from new-physics searches in the hyperon sector at BESIII is presented in Table~\ref{tab:sum}. 
In recent years, BESIII has performed a broad program of precision measurements of hyperons, creating important opportunities to probe physics beyond the SM. 
By exploiting the quantum entanglement of hyperon--antihyperon pairs, these studies demonstrate that strange baryons function as self-analyzing spin systems with exceptional sensitivity to second-generation quark dynamics, rather than merely serving as heavier analogues of nucleons.

The significance of the current results lies in their capability to test fundamental SM symmetries, particularly CP invariance and the conservation of baryon and lepton numbers. 
The landmark determination of the $\Lambda$ electric dipole moment places stringent constraints on possible CP-violating sources in the strange sector, which are closely connected to the origin of the cosmological matter--antimatter asymmetry. 
In parallel, dedicated searches for dark baryons and invisible decay modes  directly confront interpretations of the neutron lifetime puzzle, substantially restricting the viable parameter space of baryonic dark-matter scenarios and of heavy Majorana states associated with $|\Delta B| = 2$ transitions.

\begin{figure}[h]
    \centering
    \includegraphics[width=0.45\textwidth]{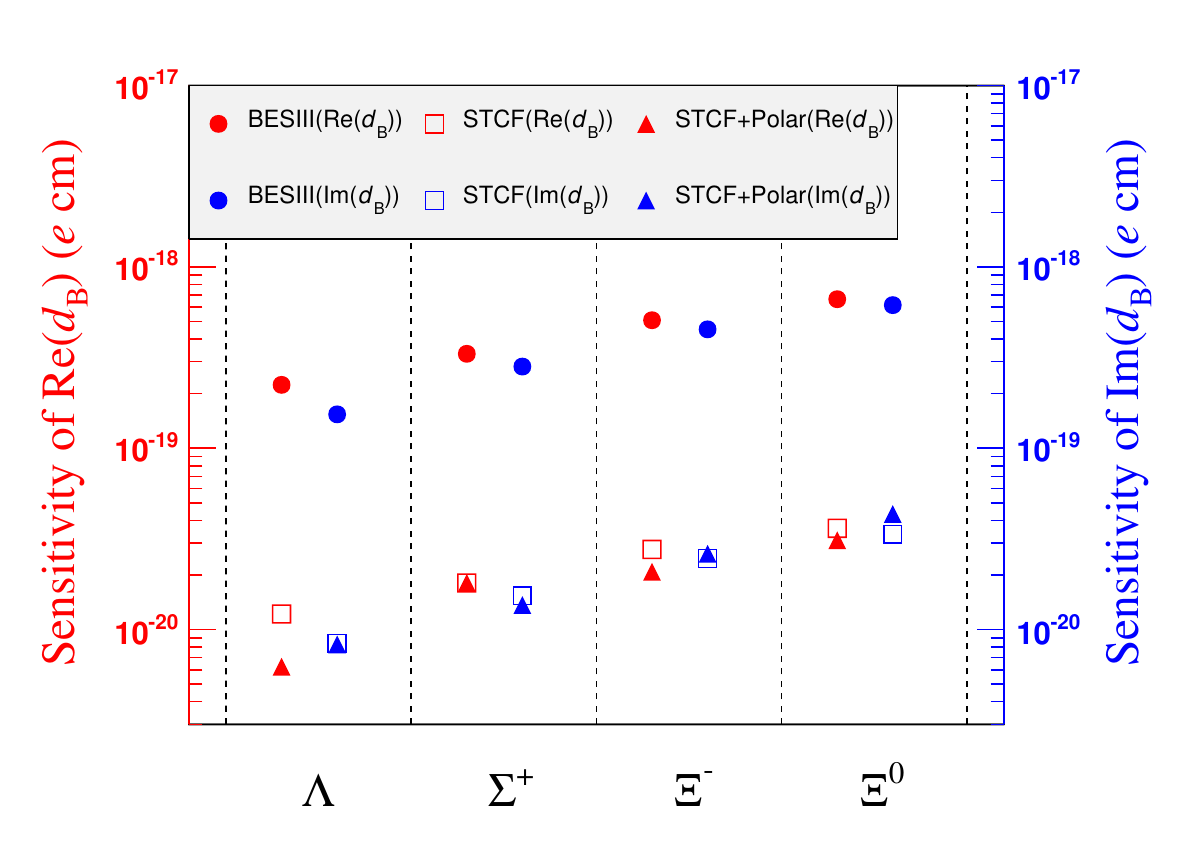}
    \caption{EDM sensitivity for the hyperons $\Lambda$, $\Sigma^{+}$, $\Xi^{-}$, and $\Xi^{0}$. 
Red and blue markers denote the observables associated with the left and right 
$y$-axes, respectively. 
}
    \label{fig:edm}
\end{figure}

Future progress in hyperon physics will require overcoming the emerging precision bottleneck driven by hadronic uncertainties and detector-related systematics. 
Key priorities include improved antineutron reconstruction together with the development of data-driven Monte Carlo and background-modeling strategies capable of keeping pace with rapidly increasing statistical samples. 
At the same time, achieving EDM sensitivities at the $10^{-19}\,e\cdot\mathrm{cm}$ level for $\Sigma^{+}$, $\Xi^{-}$, and $\Xi^{0}$ hyperons would constitute a landmark milestone: the first-ever EDM determination for these states, and in the latter two cases the first probe of hyperons containing two strange valence quarks. 
The projected measurement precisions at BESIII and at the future STCF~\cite{Fu:2023ose} are illustrated in Fig.~\ref{fig:edm}. 
Extending experimental coverage beyond the $\Lambda$ and $\Xi$ systems to higher-mass states such as the $\Omega^{-}$ is equally important, since their larger spin ($3/2$) and enhanced strangeness provide access to richer polarization observables and increased sensitivity to tensor-type BSM interactions.

The outlook for the field is particularly promising with the emergence of next-generation facilities. 
The proposed STCF, offering a luminosity roughly two orders of magnitude beyond BEPCII, is expected to accumulate trillions of $J/\psi$ events. 
Projected sensitivities indicate that EDM measurements could reach the $10^{-20}-10^{-21}\,e\cdot\mathrm{cm}$ scale---entering the region predicted by several BSM scenarios---while searches for invisible decays and rare FCNC processes may probe branching fractions down to $10^{-6}$. 
Together with complementary studies of hyperon production and dynamics at the PANDA experiment in $p\bar{p}$ collisions, these advances ensure that hyperons will remain powerful probes of hidden sectors and fundamental symmetries in the subatomic world.



\begin{acknowledgments}
We thank Professor Xiaogang He for very useful discussion. This work was supported in part by the National Key R\&D Program of China (Grant No. 2023YFA1606000), the Polish National Science Centre (Grant No. 2024/53/B/ST2/00975), and the Fundamental Research Funds for the Central Universities, University of Chinese Academy of Sciences.
\end{acknowledgments}

\bibliography{hyperon}
\end{document}